\documentclass[aps,prstab,reprint,groupedaddress,showpacs]{revtex4-1}

\usepackage{graphicx}
\usepackage{amsmath}
\usepackage{amssymb}
\usepackage{bm}

\newcommand{\f}{\begin{equation}}
\newcommand{\ff}{\end{equation}}

\begin{document}

\title[A Numerical Simulation of Chern-Simons Inflation]{ A Numerical Simulation of Chern-Simons Inflation}

\author{David Garrison and Christopher Underwood}

\address{Physics Department, University of Houston Clear Lake,  Houston, TX 77058} 

\begin{abstract}
In this work, we present results of numerical simulations of the Chern-Simons Inflation Model proposed by Alexander, Marciano and Spergel. According to this model, inflation begins with a fermion condensate interacting with a gauge field. Crucial to the success of this mechanism is the assumption that the Chern-Simons interaction would drive energy from the initial random spectrum into a narrow band of frequencies at superhorizon scales.  In this work we numerically confirm this expectation.  These gauge fields and currents, when combined with the Friedmann equations, were broken into a system of hyperbolic equations and numerically simulated.  It was found in our simulation that, by including the effects of the chiral anomaly for the axial vector current, inflation ended satisfactorily after approximately 60 e-folds.
\end{abstract}

\maketitle

\section{Introduction and Motivation}

Our understanding of the early universe is based on the phenomenon of cosmic inflation \cite{guth}.  It is well known that inflation resolves most of the problems of the standard big-bang scenario, such as the horizon problem.  Although, there are many successes of inflation, there are hundreds of scalar field models of inflation which have proven difficult to distinguish with data.  Moreover, scalar field driven inflation is fraught with conceptual and technical issues.  Recently Chen and Wang demonstrated that the primordial power-spectrum in scalar field inflation have anomalous sensitivity to high energy physics \cite{wang}.  These issues have motivated the authors \cite{alexander} to provide an inflationary mechanism that is driven by an interaction between vector and spinor fields, otherwise known as Chern-Simons inflation.  For an overview of Chern-Simons modified gravity, see the paper by Alexander and Yunes \cite{yunes}.

 Alexander, Marciano, and Spergel propose a model in which the early universe is dominated by a gauge field that interacts with a fermion current. This interaction results in an equation of state with consistently negative pressure, the condition needed for inflation~\cite{alexander}. In this model, the gauge field begins as a random, white noise spectrum; the authors assume that this evolves into a spectrum of superhorizon modes. Even though gauge fields and currents dilute with the expansion of space, their interaction energy is found to provide enough negative pressure to fuel an exponential expansion of spacetime.  Because of the complexity of the differential equations, they were not able to show this evolution analytically.  Also, in the final published version of their paper, Alexander et al. focus on the interaction between the fermion charge density and the temporal part of the gauge field.  This is most likely because they thought the spacial part of the fermion current would not be significant enough to maintain the energy density needed for inflation.  We show that while the current density does drop off quickly with volume, it can still give rise to a significant energy density.
 
This code attempts to achieve an end to inflation by modeling the Adler-Bell-Jackiw (ABJ) chiral anomaly~\cite{Adler}, which is a small quantum mechanical violation of the conservation of axial-vector current.  This violation occurs due to the tunneling of fermions from one vacuum to another.  It is by this means that the gauge field converts to leptons during inflation.  If the current decreases enough during inflation, the negative pressure generated during inflation should dwindle and inflation should end.

Another goal of this project is to establish the robustness of the model. One criticism of inflation is that specific theories can focus too much on formalism and lack a clear connection to physical processes that actually could have occurred~\cite{kinney}. Therefore, we aim to strengthen the physical interpretation of an inflationary theory while assessing its accuracy.  For this purpose, we use a new code based on the Cactus framework to simulate the early universe.

In Section 2, we provide necessary mathematical background. In Section 3, we describe the analytical and computational methods used to simulate the specific inflationary theory discussed in Section 2~\cite{alexander}. In Section 4, we present the results of the simulation, and in Section 5, we discuss the results and future research directions. 

\section{Chern-Simons Inflation} 

Because scalar field models are widely found and difficult to distinguish from one another, Alexander, Marciano and Spergel suggest an alternate model in which a gauge field interacts with fermions in the early universe to produce an effective scalar field that generates inflation~\cite{alexander}. 

In this gauge field model, we have an energy density 
\f
\rho = \frac {E^2 + B^2} {2 a^4} + |A \cdot \mathcal{J}|
\ff
where the gauge field has both an electric field, $E \equiv \dot{A}$, and a magnetic field, $B \equiv \nabla \times A$ term. We can see that this energy density has an electromagnetic component, which scales as $a^{-4}$ , as well as a gauge field-fermion interaction. The Electromagnetic terms effectively serves as a source of perturbations in the field density which later forms the basis of structure formation but does not have a significant impact on the energy density.  As shown in~\cite{alexander}, if the second term dominates over the first term, it is possible for inflation to occur. This will give us an energy density $\rho \approx |A \cdot \mathcal{J}|$ and pressure P $\approx$ -$|A \cdot \mathcal{J}|$ , corresponding to an equation of state w = -1, which is sufficient to cause inflation. 

We will consider the equation of motion of the gauge field in this case. Its action is: 
\begin{eqnarray}
S &=& \int_{M_4} d^4 x \sqrt{-g} ( \frac {M^{2}_{P} R} {8 \pi} - \frac {1} {2} \partial_{\mu} \theta \partial^{\mu} \theta + \xi R \theta^2  \nonumber \\ &-& \frac {1}{4} Tr [ F_{\alpha \beta} F^{\alpha \beta} ] 
+ \frac {\theta}{4 M_{*}} Tr [ F_{\alpha \beta} F^{\alpha \beta} ] \nonumber \\  &+& q Tr [ A_{\mu} \mathcal{J}^{\mu}_{5}] )
\end{eqnarray}

By varying this action with respect to the gauge field, we find the equation of motion.  The gauge field$\textquoteright$s equation of motion, in terms of Fourier modes, is then ~\cite{alexander}: 
\f
\ddot{A}_h + k^2 A_h = - h k A_h \frac {\dot{\theta}} {M_{*}} + a^4 \mathcal{J}_h 
\ff 
where h refers to the different helicities,  $M_{*}$ is the mass scale identified with the UV cut-off scale of the effective field theory and $\theta$ is responsible for CP violation.  The fermion currents are 
\f
\sqrt{2} \mathcal{J}_h =  \mathcal{J}_1 + h i \mathcal{J}_2 .
\ff
Alexander et al. assume that the fermion currents have a background solution in which they scale as $J_0 / a$.  Since they are being suppressed by a factor of a, we can define a set of constants $J_h$ where $J_h = \mathcal{J}_h / a$~\cite{alexander}. The equation of motion for the gauge field is now 
\f
\ddot{A}_h + k^2 A_h = - h k A_h \frac {\dot{\theta}} {M_{*}} + a^3 J_h 
\ff 
According to the solutions for the gauge field~\cite{alexander}, a value of $k < \frac {\dot{\theta}} {M_{*}}$, corresponding to long-wavelength modes, will result in its exponential growth. However, the presence of short wavelength modes, $k > \frac {\dot{\theta}} {M_{*}}$ , will result in continued linear growth due to back-reaction of the gravitational field.  

We model the flow of current with the ABJ anomaly.  It was shown by Adler~\cite{Adler} that the axial-vector current is not conserved in quantum field theory.  This anomaly arises through perturbation theory and the analysis of triangle diagrams with a gauge-invariant regularization procedure.  During inflation, the anomaly causes the gauge field to decay into fermions.

The divergence of the axial fermion current in the ABJ anomaly is defined by:
\begin{equation}
{\nabla _\mu }{\mathcal J}_5^\mu  = \frac{1}{32\pi^2}{F_{\alpha \beta }}{F_{\mu \nu }}{\epsilon^{\alpha \beta \mu \nu }}.
\end{equation}
We assume the spacial components of the current act like a charge density, $J^0$, which moves with a velocity, $v^i$:
\begin{equation}
J^i = J^0 v^i.
\end{equation}
For this simulation, we assume the velocity to be constant and only model the changes in current through the charge density.  We therefore derive the time rate of change of the charge density for the ABJ anomaly to be:
\begin{eqnarray}
{\partial _0}J^0 =  &&\frac{\vec E \cdot \vec B}{4{\pi ^2}{a^2}} - {\partial _i}{J^i} - 2HJ^0.
\end{eqnarray}
In order to maintain an initial current density of approximately $10^{-20} M_p^3$, the initial charge density and current velocity were chosen to be:
\begin{equation}
J^0 = 10^{-10}M_p^3  ~~~\mbox{and}~~~  v^i = 10^{-10}.
\end{equation}
The choice of these two numbers is arbitrary as long as the velocity is not relativistic and that they multiply to something on the order of $10^{-20}M_p^3$.

The $\theta$ field is determined by:
\begin{equation}
\ddot \theta  + 3H\dot \theta  + 2{m^2}\theta  = \frac{{\vec E \cdot \vec B}}{{4{a^3}{M_*}}},
\end{equation}
where $m$ is on the order of the GUT energy scale.  Assuming that $m \approx H$ and that the right hand side of (10) becomes negligible during inflation, the solution to (10) is $\theta (t) = -{\theta_0}e^{-Ht}$ \cite{alexander}.  Our simulation does not rely on this assumption, we allow $\theta$ to be determined dynamically.

Because the universe was essentially a plasma before last scattering, we look to plasma physics for an analogy of how our system should behave. Specifically, we consider magnetohydrodynamics (MHD), the study of electrically conducting fluids.  Here in the Chern-Simons system, the gauge field is considered analogous to the field part of the MHD equations while the velocity terms are considered constant and the density is free to vary like in a compressible MHD system.  When fluids in motion are electrically conducting, currents in the fluid produce magnetic fields that induce forces and thus change the dynamics of the system when it becomes turbulent~\cite{shebalin}. This kind of turbulence is analogous to the dynamical system changes that could be capable of generating inflationary behavior. Computer codes that simulate MHD use the equations of fluid dynamics, along with Maxwell$\textquoteright$s equations, to analyze the dynamics of a plasma system. Simulations such as these are commonly known to lead to what is referred to as an ``inverse energy cascade''.  In a hydrodynamic turbulent system, as a system evolves in time, energy tends to ``cascade'' from the longest wavelength modes to the shortest wavelength modes as turbulent eddies get smaller and smaller.  In MHD systems, the reverse happens.  Energy cascades inversely from high frequency modes to low frequency modes.  

Utilizing the coupled partial differential equations that describe the very early universe could allow us to see the resulting exponential increase and then ``leveling off'' of the scale factor. Additionally, a simulation of this type will allow us to see directly the physical effects of fine-tuning the initial conditions for these coupled differential equations. We can input the equations of motion for the fields and currents that are interacting to generate inflation, along with the Friedmann equations, which describe the expansion of space, in order to create a system similar to the MHD equations. The Friedmann equations will allow us to relate the evolving pressure and density of the system to the evolution of the scale factor. 

\section{Methods and Equations}

For the numerical calculation, we use natural units but later evaluate the data in terms of SI units so that the results can be easily compared to the established values. In this simulation, we allow the gauge field to vary so the most relevant equation from~\cite{alexander} is the equation of motion for the gauge field, A: 
\f
\ddot{A} -  \nabla^2 A + \frac {\dot{\theta}} {M_{*}}  (\nabla \times A) - a^4 \mathcal{J}_i  = 0 
\ff
In comoving coordinates this equation becomes:
\f
a(t) \frac {d^2 A} {dt^2} + \frac {da} {dt} \frac {d A} {dt} = J a(t)^3 + \frac {1}{a(t)} \nabla^2 A -  a(t)^2 \frac {\dot{\theta}} {M_{*}} \nabla \times A
\ff
In order to use this in our code, we separated the equations of motion for the gauge field, ABJ chiral anomaly, Chern-Simons term and the Friedman equations into a system of first order in time differential equations.  
\begin{eqnarray}
\frac {d A} {dt} = && \frac {Z} {a} \\
\frac {d Z} {dt} = && J a^3 + \nabla^2 A / a - a^2  \frac {\dot{\theta}} {M_{*}}  (\nabla \times A) \\
\frac {d J^0} {dt} =  && \frac{\vec E \cdot \vec B}{4{\pi ^2}{a^2}} - {\partial _i}{J^i} - 2H J^0 \\
\frac{d D}{dt} = && \frac{\vec E \cdot \vec B}{4a^3 M_*^2} - 3 H D - 2 \frac{m^2} {M_*} \theta \\
\frac{d\theta}{dt} = && D M_* \\
\frac {d a} {dt} = && a H \\
\frac {d H} {dt} = && \frac {8 \pi} {3} \bar{\rho} / a - H^2
\end{eqnarray}
The average energy density and pressure are calculated as $\bar{\rho} = \frac{1}{N} \sum_{k=1}^N |A_k \cdot \mathcal{J}_k|$ and $\bar{P} = -\frac{1}{N} \sum_{k=1}^N |A_k \cdot \mathcal{J}_k|$.  Here N represents the total number of grid points in the computational domain.  The scale factor and Hubble parameter therefore depend on the average energy density and not the local field dynamics.  
D = $\frac {\dot{\theta}} {M_{*}}$ is calculated at each timestep.

The gauge field was composed of a random (white noise) spectrum.  In order to generate the initial gauge field, we used a random number generator to create a random spectrum with amplitude up to the calculated maximum amplitude, $A_0$, in each direction.  The magnitude of the gauge field was then held equal to the initial amplitude $A_0$.  The initial value for the other variables is given below.
\begin{eqnarray}
| A_0 | = && 3.36566 \times 10^{-5}M_{P} \\
J^0 = && 10^{-10} M_{P}^3 \\
\vec{v} = && 10^{-10} (\hat{x} + \hat{y} + \hat{z}) \\
\frac {\dot{\theta}_0} {M_{*}} = && 2.18 \times 10^{-5} M_{P}  \\
a_0 = && 1.0 \\
H_0 = && \sqrt{\frac{8 \pi} {3} | A_0 \cdot J |} \\
Z_0 = && H_0 a_0 \times random~number(-1,1) \\
m = && 4.0 \times 10^{-6}M_{P} \\
M_* = && 4.0 \times 10^{-6}M_{P} 
\end{eqnarray}
The code was then run on the University of Houston Texas Learning \& Computation Center$\textquoteright$s Xanadu cluster using a variety of time-steps, grid sizes and resolutions in order to obtain consistent results.   We ran the code using resolutions ranging from $34^3$ to $132^3$ grid points and grid sizes ranging from $10^4$ to $10^7$ units cubed.  These grid sizes correspond to the size of the horizon throughout the simulation.  We adjusted the Courant ratio from a low of 0.001 to a high of 0.9.  These runs used anywhere from 8 to 48 processors and the runs lasted anywhere from 2 hours to 7 days.  We also ran the code using three different differencing schemes including 2nd order finite differencing, 4th order finite differencing and Fourier spectral differencing.   
	
Because the initial units were entered as Planck units, we assumed that the physical grid (horizon) size corresponded to Planck lengths and the timing output could be interpreted as Planck time.  The output was analyzed using several tools including ygraph and Pro Fit (www.quansoft.com). 

\section{Results}

Using the definition that inflation is active during regions where $\ddot a >0$, we saw that inflation consistently began at around $2.0 \times 10^{-37}$s and lasted until $1.5 \times 10^{-36}$s before settling into a linear progression as shown in Fig. 1.  The values for $m$ and $M_*$ that produced the best results was $4.0 \times 10^{-6}M_p$ for both parameters.  In order to achieve at least 60 e-folds of inflation, the initial gauge field parameter, $A_0$, necessary was $3.36566 \times 10^{-5}M_p$.  The code was highly sensitive to this number and even a deviation of $10^{-10}M_p$ in either direction caused the scale factor to go infinite or level off too soon. These parameters were tuned using a grid size of $10^6$ Planck lengths. When the code was run using these parameters on a grid size of $10^5$ Planck lengths no inflation occurred at all, and when run using a grid size of $10^7$ Planck lengths the scale factor leveled off at about $10^3$ most likely because of the reduced effective resolution for the larger grid size.

\begin{figure*}
\includegraphics[height=3.9in,width=7.0in,angle=0]{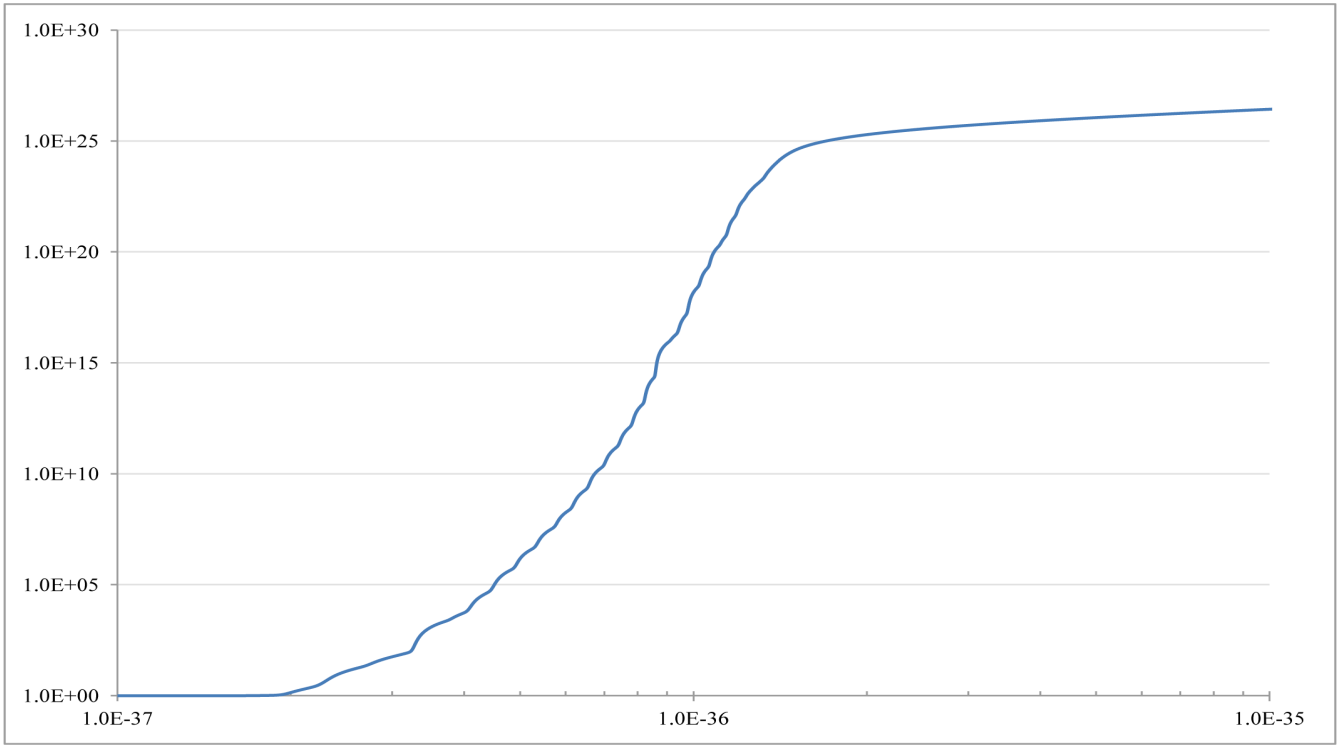}
\caption{The scale factor for the inflationary period (log-log scale).\label{fig:scale}} 
\end{figure*}
\begin{figure*}
\includegraphics[height=2.6in,width=7.0in,angle=0]{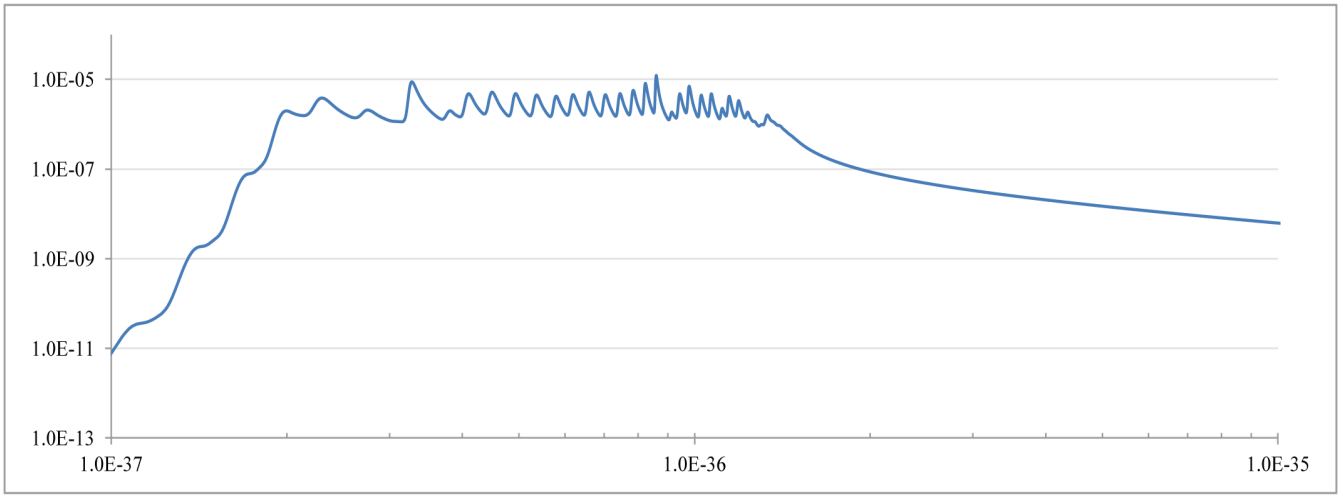}
\caption{The Hubble parameter throughout the inflationary period (log-log scale).\label{fig:hubble}} 
\end{figure*}
\begin{figure*}
\includegraphics[height=2.0in,width=3.5in,angle=0]{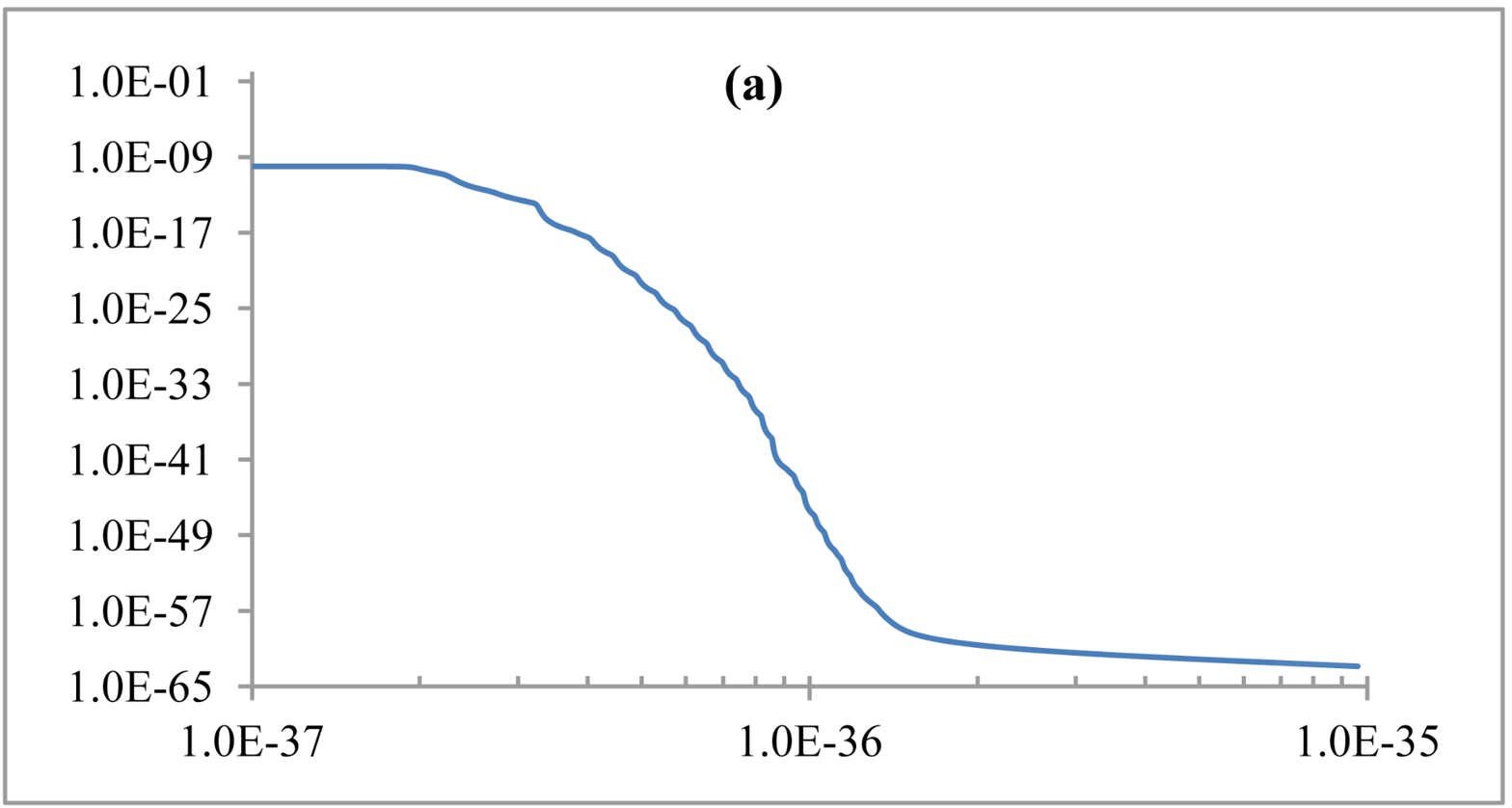}
\includegraphics[height=2.0in,width=3.5in,angle=0]{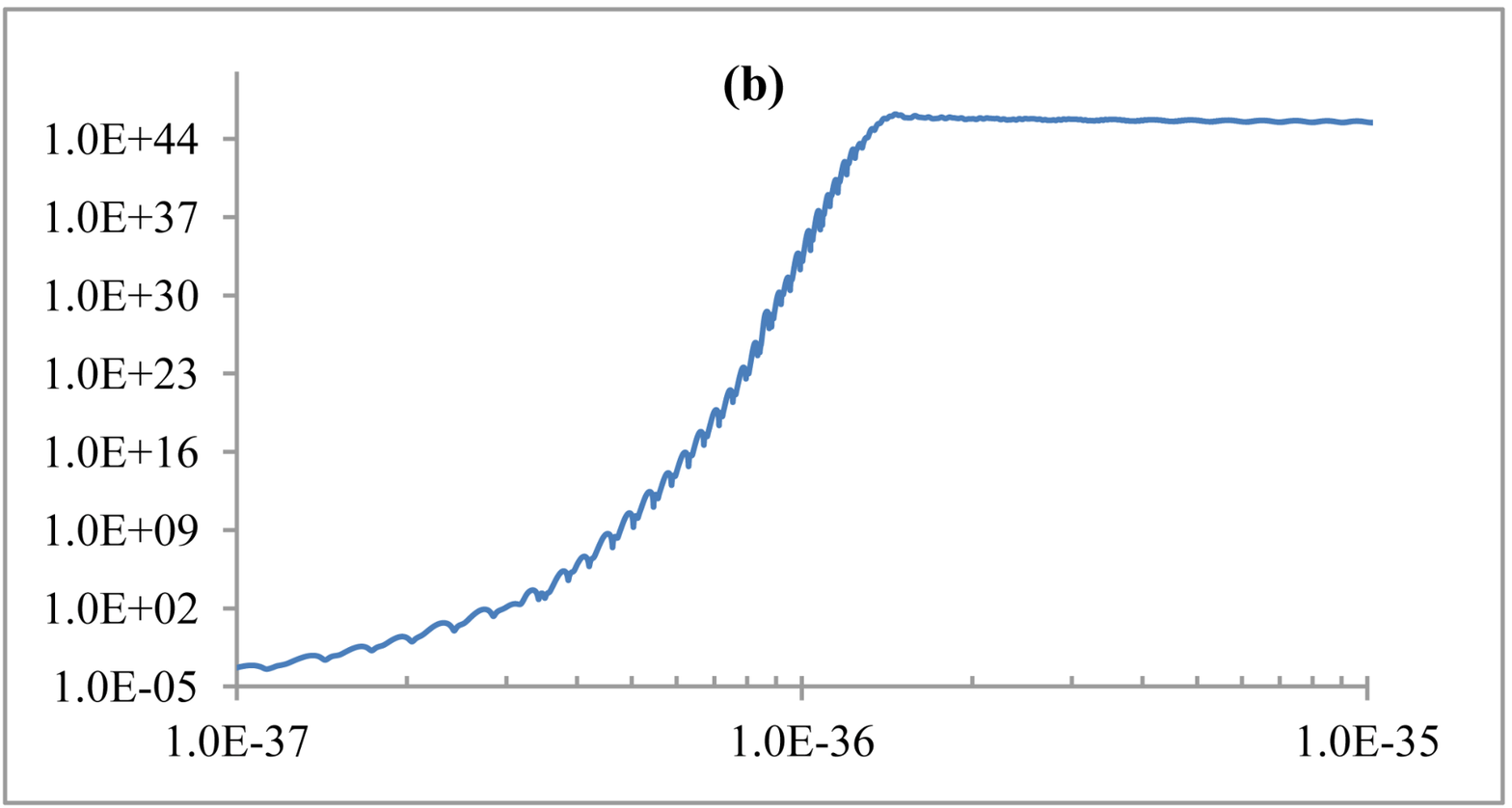}
\caption{The (a) charge density and (b) magnitude of the gauge field throughout inflation (log-log scales).\label{fig:J0}} 
\end{figure*}

The Hubble parameter in Fig. 2 rose sharply from $1.0 \times 10^{-37}$ to $2.0 \times 10^{-37}$s to about $2.0 \times 10^{-6}M_p$ before undergoing an oscillatory behavior.  The oscillations increased in frequency and persisted until $1.4 \times 10^{-36}$s where it settled into a exponential decline.  These oscillations seem to be the result of oscillations in the $\theta$ term.  The behavior of the charge density in Fig. 3(a) is the opposite of that of the scale factor.  It reached as low as $10^{-60}M_p^3$ before leveling off.  Notice that although the charge density drops of quickly, as $a^{-3}$, it still has a significant impact on the  energy density because of the high growth rate of the gauge field.  The magnitude of the gauge field (Fig. 3(b)) rose sharply during inflation before leveling off at around $10^{50}M_p$.  This implies that the gauge field grew like $a^{2}$ not as $a^{1}$ throughout most of the inflation period.  There also appears to be a period at the beginning of the simulation when the gauge field grew as fast or faster than the charge density declined, resulting in a constant or growing energy density and therefore a constant or growing Hubble Parameter.  Because of the variations in the growth rates of the gauge field and fermion current, the inflation event ends naturally instead of continuing indefinitely.

\begin{figure*}
\includegraphics[height=1.5in,width=3.5in,angle=0]{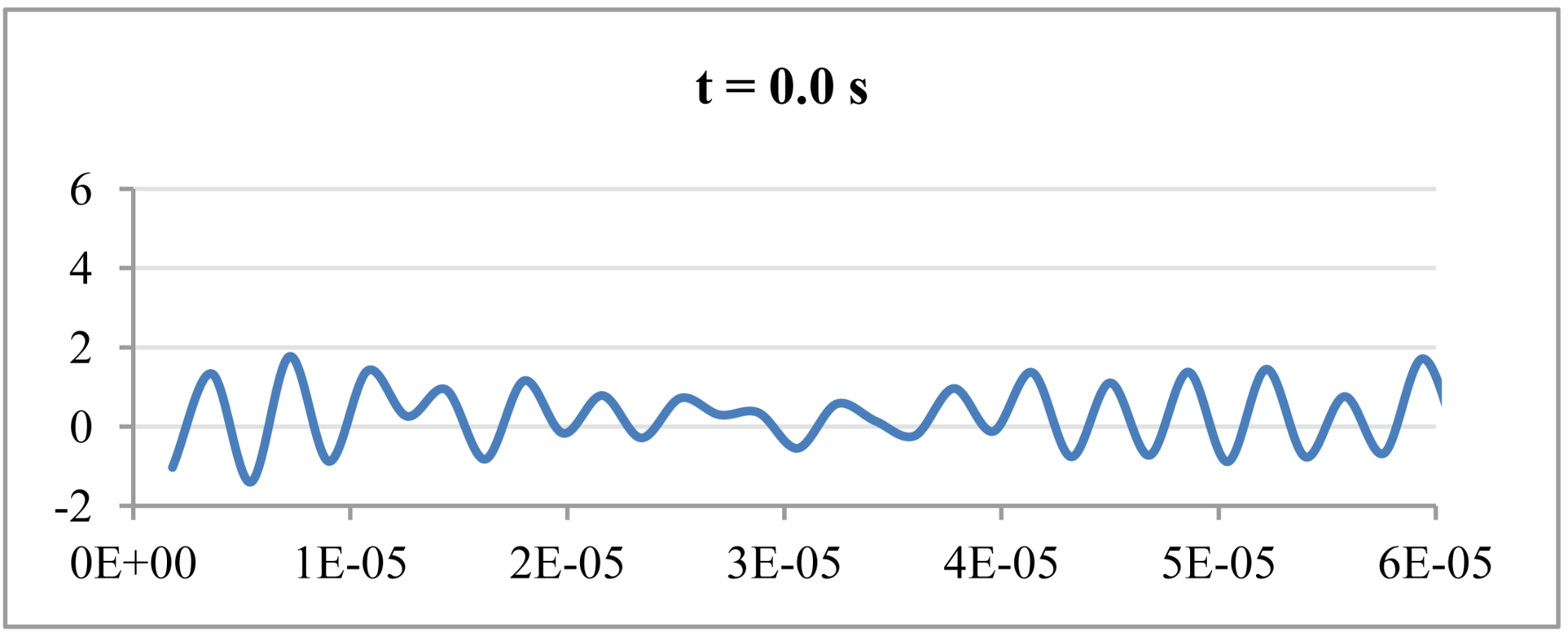}
\includegraphics[height=1.5in,width=3.5in,angle=0]{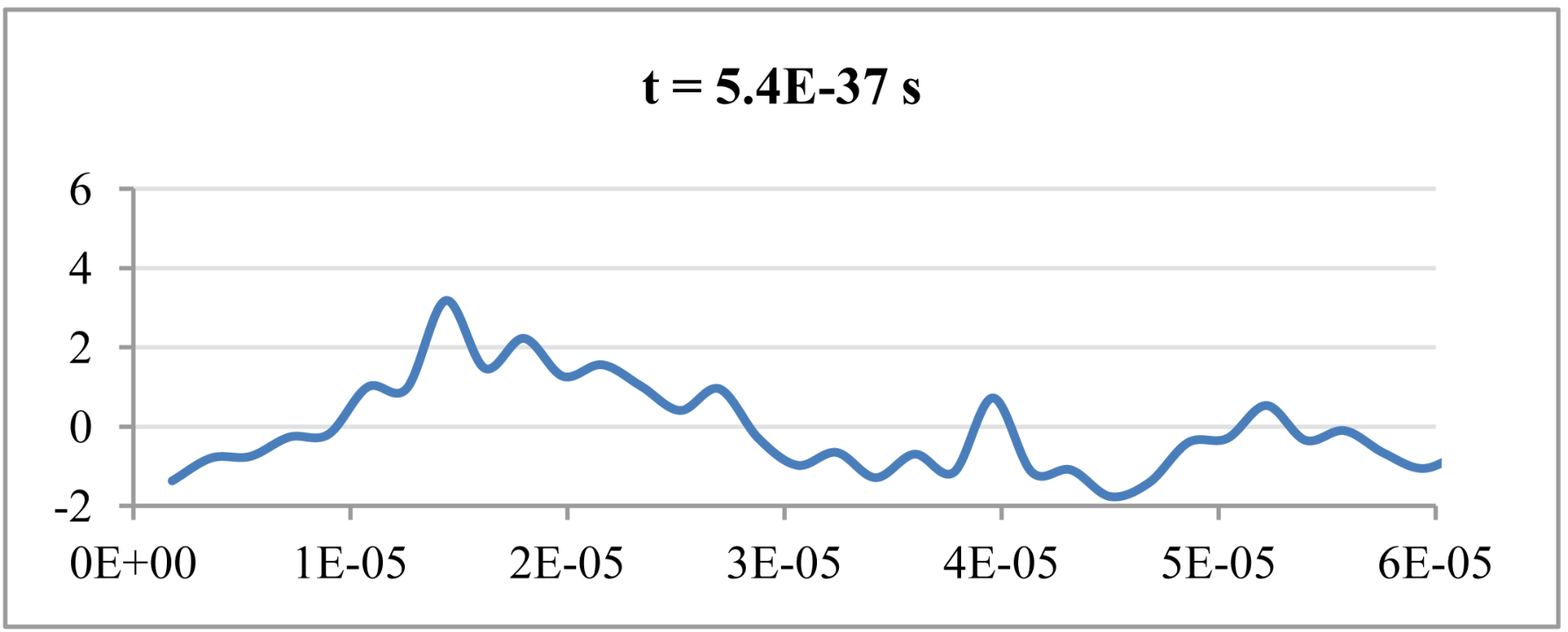}
\includegraphics[height=1.5in,width=3.5in,angle=0]{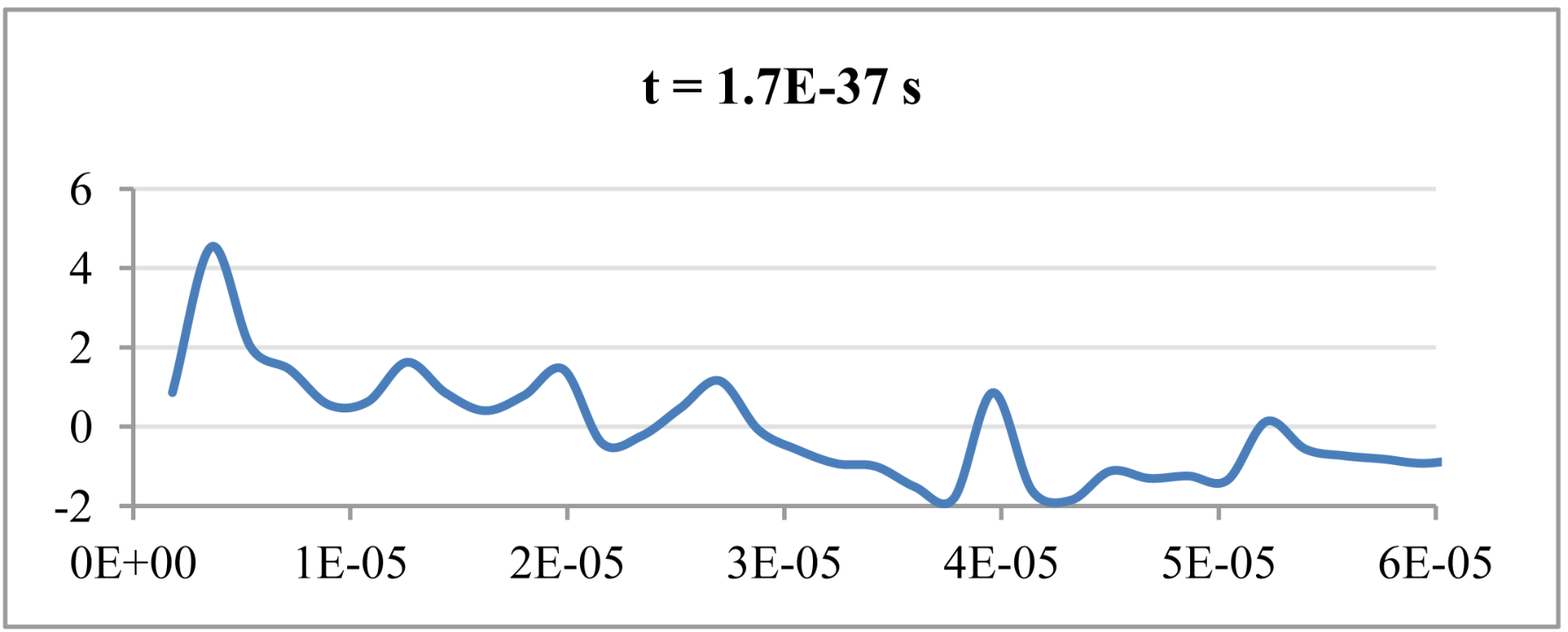}
\includegraphics[height=1.5in,width=3.5in,angle=0]{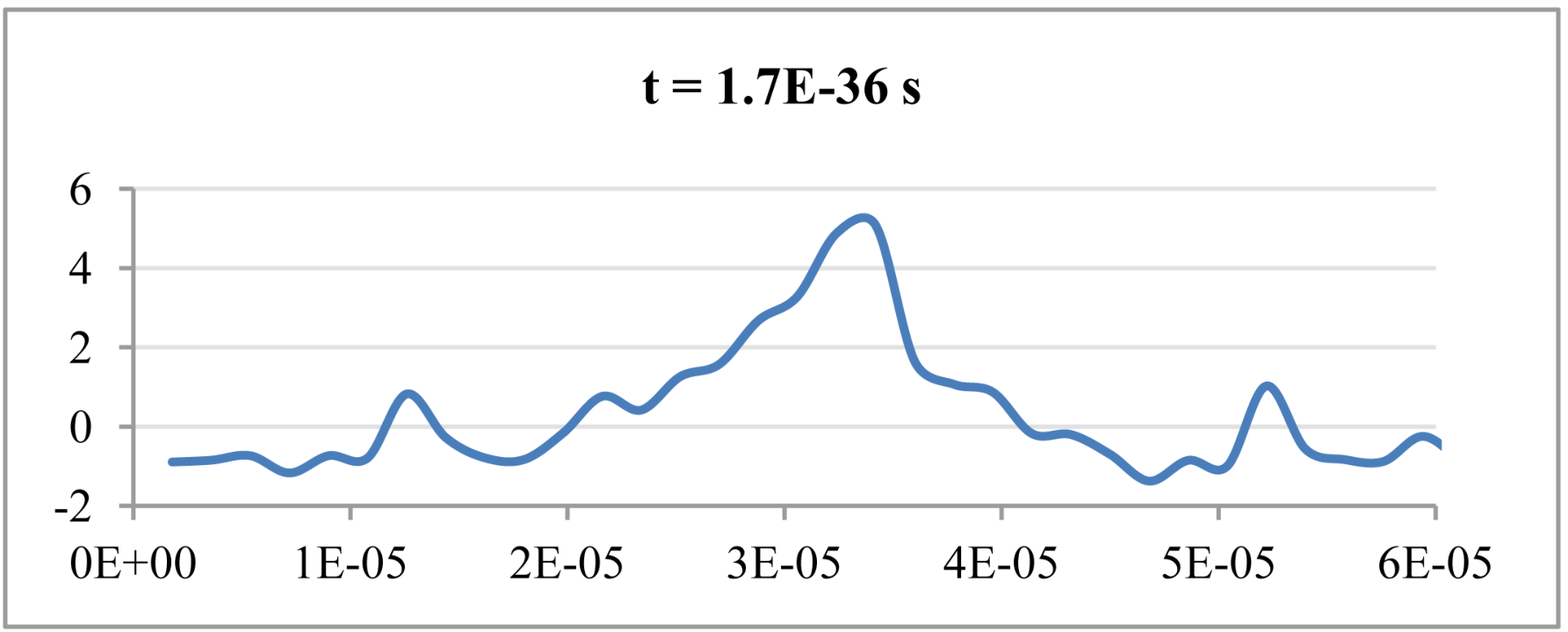}
\caption{Power spectral density for the gauge field throughout inflation.  The vertical axis is the log of the deviation from the mean.  The horizontal axis is frequency in Plank units.\label{fig:psd}} 
\end{figure*}

The power spectrum (Fig. 4) of the gauge field is initially random, but flattens out before the onset of inflation, with lower frequency modes becoming dominant.  In the middle of the inflationary period, the spectrum starts to become concentrated in higher frequency modes before accumulating into wide band of mid-range frequencies from $2.0 \times 10^{-5}$ to $5.0 \times 10^{-5}t_p^{-1}$ at the end of inflation.  This wide band power spectrum persists throughout the post-inflation era.

\section{Discussion and Conclusions}

The simulation presented here clearly supports much of the hypotheses proposed by Alexander, Marciano and Spergel: the evolution equations given drive energy into a narrow band of modes and cause the universe to expand exponentially. Though the theory does not predict an initial start time for inflation, our value of about $t_i = 10^{-37}$ s is consistent with the time predicted by the most general models of inflation~\cite{weinberg}. This suggests some universality in the initial conditions required for inflation, regardless of the model; in the future, we can study this question analytically for this model. 

The power spectrum in Fig. 4 supports the hypothesis that an initially random gauge field spectrum becomes concentrated in low frequencies during inflation. This analysis allows us to see the progression from an almost completely random spectrum near the beginning to the energy being concentrated in a narrow band of low-frequency modes during the inflation event. In addition to the dominance of low-frequency modes, the higher modes are suppressed during much of inflation. This reflects the behavior predicted in~\cite{alexander}. The dominance of low-frequency modes has subsided, and we are mostly left with lower amplitude random noise behind a flatter power spectrum, indicating that the end of inflation is near. 

It is interesting that exponential growth does not occur for horizon sizes that are too small. This could be due to the unavailability of very long-wavelength (low frequency) modes for the gauge field. In the future, accurate measurements of these frequencies could allow us to verify that they are in fact, superhorizon modes. A more detailed study of the effect of grid resolution on the dynamics of the system could lead to some interesting revelations about the role of low and high frequency modes in the system.  By understanding the spectral dependence on the dynamics of the system we can better understand how our choice of initial conditions effects the system.  We can then better determine what initial conditions lead to the universe which we observe today.  Other future work on this project will be to incorporate the product of the charge density and temporal part of the gauge field into the energy density to determine if they also have a significant impact on inflation and the dynamics of the system.  

The results presented here only work for the grid size and grid resolutions discussed.  This speaks to the level of fine tuning necessary in this model to get the required 60 e-folds of inflation.  The value of $4.0 \times 10^{-6}M_p$ for $m$ is approximately the GUT scale for string theory.  Increasing the initial gauge field had by far the most effect of the scale factor.  Since the plot of the charge density closely resembles a mirror image of the scale factor, the Hubble parameter term in (8) is likely responsible for most of the decrease in charge density, as well as energy density, and the anomalous behavior of the axial current likely had little effect.  Typical models have inflation taking place from approximately $10^{-37}$ to $10^{-32}$s.  The inflation described by this simulation starts at about $10^{-37}$s and ends at roughly $10^{-36}$s which is far quicker than what most models predict.  This may be resolved by the right choice of initial parameters with sufficiently high resolution.

\begin{acknowledgments}
Thanks to the University of Houston's Texas Learning and Computing Center for access to and use of the Xanadu computing cluster. Thanks to Stephon Alexander and Annie Preston for many useful conversations.
\end{acknowledgments}

\section*{References}

\bibliography{bibliography}

\end{document}